\documentclass[a4paper,fleqn]{cas-dc}

\usepackage[numbers]{natbib}

\usepackage{graphicx}
\usepackage{amssymb}
\usepackage[caption=false]{subfig}
\usepackage{multirow}
\usepackage{dcolumn}
\usepackage{bm}
\usepackage{threeparttable}
\usepackage{booktabs}
\usepackage{color}
\usepackage{booktabs}
\usepackage{xspace}
\usepackage{chemformula}
\usepackage{xcolor}
\usepackage{lipsum}
\usepackage[colorinlistoftodos]{todonotes}
\presetkeys{todonotes}{inline}{}
\usepackage{siunitx}
\usepackage[capitalise]{cleveref}
\usepackage{amsmath}
\usepackage{lineno}

\newcommand{\CTO}{CoTiO$_{3}$\xspace}
\newcommand{\cm}{cm$^{-1}$\xspace}

\def\tsc#1{\csdef{#1}{\textsc{\lowercase{#1}}\xspace}}
\tsc{WGM}
\tsc{QE}
\tsc{EP}
\tsc{PMS}
\tsc{BEC}
\tsc{DE}

\begin{document}
\let\WriteBookmarks\relax
\def\floatpagepagefraction{1}
\shorttitle{Lattice dynamics and spontaneous magnetodielectric effect in ilmenite CoTiO$_{3}$}
\shortauthors{R.~M.~Dubrovin et~al.}

\title [mode = title]{Lattice dynamics and spontaneous magnetodielectric effect in ilmenite CoTiO$_{3}$}                      



\author[Ioffe]{R.~M.~Dubrovin}[orcid=0000-0002-7235-7805]
\cormark[1]
\cortext[cor1]{Corresponding author}
\ead{dubrovin@mail.ioffe.ru}
\credit{Conceptualization, Validation, Formal analysis, Investigation, Writing - Original Draft, Visualization}

\author[Ioffe]{N.~V.~Siverin}[orcid=0000-0002-4643-845X]
\credit{Investigation, Validation, Formal analysis, Visualization}


\author[HFML,Radboud]{M.~A.~Prosnikov}[orcid=0000-0002-7107-570X]
\credit{Validation, Formal analysis, Investigation, Writing - Original Draft, Visualization}

\author[URFU]{V.~A.~Chernyshev}[orcid=0000-0002-3106-3069]
\credit{Formal analysis}

\author[ISAN]{N.~N.~Novikova}[orcid=0000-0003-2428-6114]
\credit{Investigation}

\author[HFML,Radboud]{P.~C.~M.~Christianen}
\credit{Resources}

\author[MPEI]{A.~M.~Balbashov}
\credit{Resources}

\author[Ioffe]{R.~V.~Pisarev}[orcid=0000-0002-2008-9335]
\credit{Writing - Review \& Editing, Supervision, Project administration, Funding acquisition}

\address[Ioffe]{Ioffe Institute, Russian Academy of Sciences, 194021 St.-Petersburg, Russia}
\address[HFML]{High Field Magnet Laboratory (HFML - EMFL), Radboud University, Toernooiveld 7, 6525 ED Nijmegen, The
Netherlands}
\address[Radboud]{Radboud University, Institute for Molecules and Materials, Heyendaalseweg 135, 6525 AJ Nijmegen, The
Netherlands}
\address[URFU]{Department of Basic and Applied Physics, Ural Federal University, 620002 Ekaterinburg, Russia}
\address[ISAN]{Institute of Spectroscopy, Russian Academy of Sciences, 108840 Moscow, Troitsk, Russia}
\address[MPEI]{Moscow Power Engineering Institute, 111250 Moscow, Russia}

\begin{abstract}
Ilmenite-type crystals find a variety of technological applications due to their intriguing physical properties.
We present the results of the lattice dynamics studies of honeycomb antiferromagnetic ilmenite \CTO single crystal by the complementary polarized infrared, Raman, and dielectric spectroscopic techniques that are supplemented by the DFT calculations.
The frequencies and symmetries of all predicted infrared and Raman active phonons were uniquely identified.
Furthermore, it was found that the dielectric permittivity demonstrates distinct changes below antiferromagnetic ordering temperature in zero magnetic field due to spontaneous magnetodielectric effect.
Our results establish the reliable basis for further investigation of the coupling of phonons with spins, magnetic excitations and other physical phenomena of this promising material.
\end{abstract}



\begin{keywords}
CoTiO$_{3}$ \sep Ilmenite \sep Single crystals \sep Lattice dynamics \sep Infrared spectroscopy \sep Raman spectroscopy \sep Dielectric spectroscopy \sep DFT calculations
\end{keywords}

\maketitle

\section{Introduction}

Titanate materials \ch{ATiO3} have an astonishing variety of different crystal structures depending on the chemical composition, which is manifested in numerous intriguing physical phenomena such as ferroelectricity, magnetism, multiferroicity, piezoelectricity, and some others~\cite{tilley2016perovskites}.
Among them, cobalt titanate \CTO, which possesses a very stable ilmenite structure in a wide range of temperatures and pressures~\cite{cunha2019thermal}, has numerous successful industrial applications such as high-$\kappa$ dielectric~\cite{chao2004cotio}, resonator antenna~\cite{ullah2015design}, catalysts~\cite{hwang2018thermogravimetric,agafonov2008catalytically,ehsan2019fabrication}, gas sensors~\cite{siemons2007gas,lu2018rgo}, a potential anode material for \ch{Li}-ion batteries~\cite{xu2017situ,li2017ternary,li2019mof,jiang2013diffusion}.
From the condensed matter physics point of view the \CTO exhibits exciting phenomena, e.g., Dirac magnons~\cite{yuan2020dirac,elliot2020visualization} and magnetodielectric coupling~\cite{harada2016magnetodielectric} whereas isostructural \ch{MnTiO3} reveals linear magnetoelectric effect~\cite{mufti2011magnetoelectric,silverstein2016incommensurate} and magnetochiral dichroism~\cite{sato2020magnetochiral}.
But further intensive research of this promising material is inhibited by the lack of complete information about the phonon spectrum.
It is worth noting that up to now, most experimental studies of the lattice dynamics of \ch{ATiO3} materials with ilmenite structure were performed on the poly- and microcrystalline samples using unpolarized far infrared~\cite{baran1978ir,baran1981infrarotspectrum,yamaguchi1987formation,hofmeister1993ir,wang2008far,rodrigues2020unveiling} and Raman~\cite{baran1979raman,ko1989phase,baraton1994vibrational,wang2008assignment,wu2011investigation,fujioka2016raman,sakai2017raman,rodrigues2018first} spectroscopy.

In this paper, we present results on the lattice dynamics study of \CTO single crystal with ilmenite structure using complementary dielectric, far-infrared reflectivity, and Raman scattering polarized spectroscopic techniques.
Obtained experimental results, supplemented by the lattice dynamics calculations, allowed us to determine the frequencies and symmetries of infrared and Raman active phonons.
Moreover, we show that the antiferromagnetic ordering leads to notable changes in the dielectric permittivity in zero magnetic field due to the spontaneous magnetodielectric effect.

\section{Experimental and Computational Details}

The \CTO single crystal was grown by floating zone melting method with light heating in oxygen flow using polycrystalline powder synthesized with \ch{Co3O4} and \ch{TiO2} of 4N purity as describes in details in Ref.~\cite{balbashov2017electric}.
The X-ray oriented single crystal was cut close to normal to the hexagonal $a$ and $c$ axes and optically polished.
The far-infrared reflectivity measurements were carried out using Fourier-transform IR spectrometer Bruker IFS 66v/S in the spectral range of 50--7500\,\cm at room temperature.
Raman scattering spectra were measured in the range 15--1200\,\cm with the use of FHR1000 (HORIBA) monochromator equipped by 1200 lines/mm grating, \SI{100}{\um} entrance slit, liquid nitrogen cooled PyLoN CCD camera (Princeton Instruments) and \SI{660}{\nm} excitation laser (Torus, LaserQuantum).
Low excitation power of \SI{500}{\uW} was used to avoid the overheating of an optically dense sample.
Electric measurements of dielectric permittivity were done using precision RLC meter AKTAKOM AM-3028 in the frequency range from \SI{20}{\Hz} to \SI{1}{\MHz}.
Electric contacts were deposited on the sample faces using silver paint to form a capacitor.
Measurements were performed in helium flow cryostat Cryo CRC 102 in the temperature range from 5 to 400\,K at continuous heating.

The measurements were complemented by lattice dynamical simulations which were performed according to the density functional theory (DFT) with the \textsc{B3LYP} hybrid functional~\cite{becke1993density} implemented on \textsc{CRYSTAL14} package~\cite{dovesi2014crystal14}.
All-electron basis sets TZVP~\cite{peintinger2013consistent} for \ch{Co}, 8-6411(31d)G~\cite{sophia2013first} for \ch{Ti}, and 8-411(1d)G~\cite{sophia2013first} for \ch{O} have been used according to Ref.~\cite{rodrigues2020unveiling}.
The reciprocal space integration was performed by sampling the Brillouin zone with the $12 \times 12 \times 12$ Pack-Monkhorst $k$ mesh.
Coulomb and exchange integral tolerance factors were set to tight values of $10^{-8}$, $10^{-8}$, $10^{-8}$, $10^{-9}$, and $10^{-30}$.
The optimization of crystal structure was carried out and the experimental lattice parameters from Ref.~\cite{balbashov2017electric} were well reproduced with an error of about 1.2\%.
The phonon spectrum was calculated by means of numerical second derivatives of the total energy as described in Ref.~\cite{kuzmin2020lattice}.
The static dielectric tensor, Born effective charges and phonon intensities were calculated using the \textsc{CPHF/KS} approach~\cite{maschio2012infrared,maschio2013raman}.

\section{Results and Discussion} \label{S:Results}

\begin{figure}
\centering
\includegraphics[width=\columnwidth]{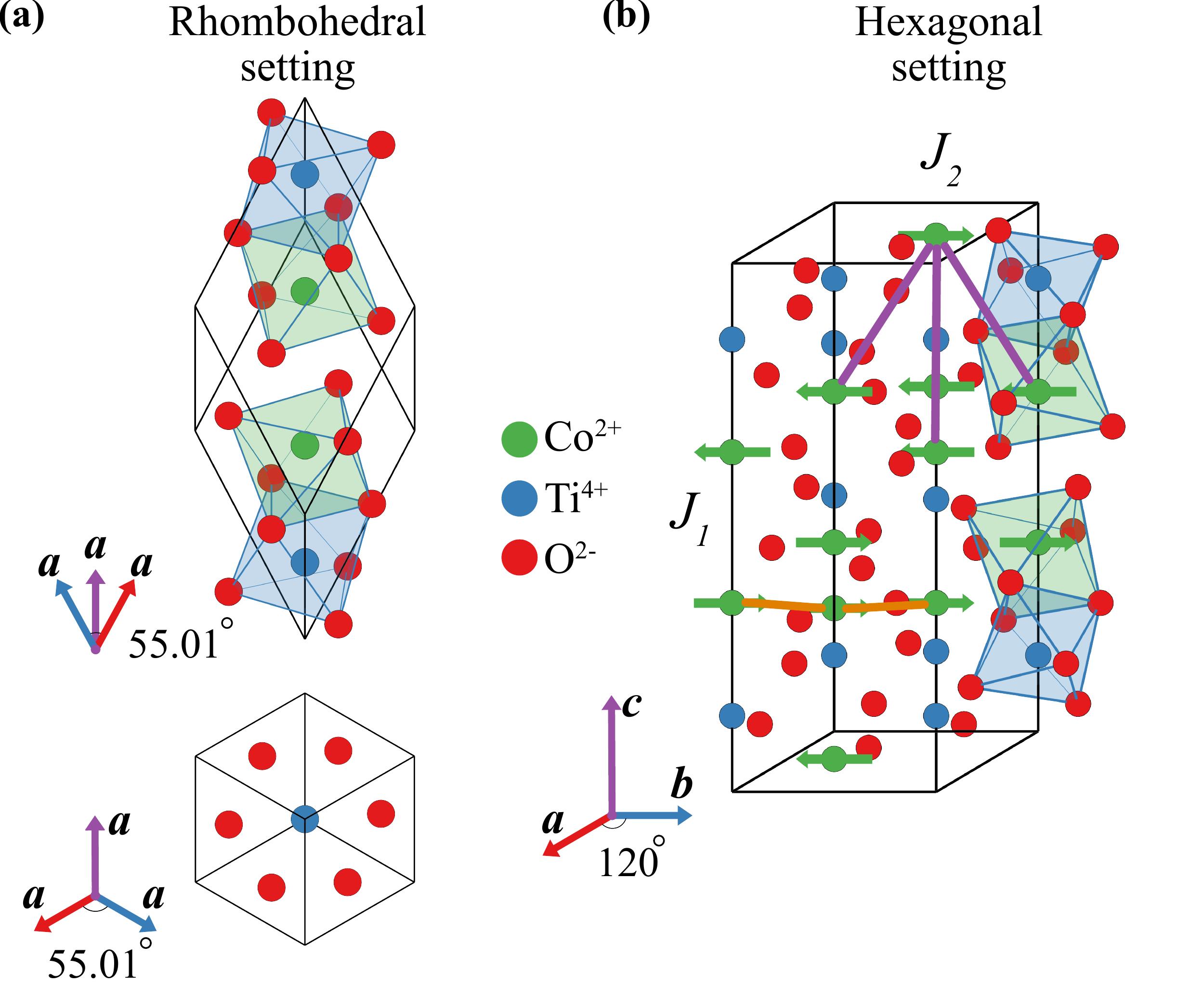}
\caption{\label{fig:CoTiO3_structure}
Crystal structure of ilmenite \CTO with trigonal space group $R\overline{3}$ in the (a)~rhombohedral and (b)~hexagonal settings.
The orange and purple lines indicate the principal superexchange interaction $J_{1}$ and $J_{2}$, respectively.
Pictures were prepared using the \textsc{VESTA} software~\cite{momma2011vesta}.
}
\end{figure}

\CTO has an ilmenite crystal structure with trigonal space group $R\overline{3}$ ($C^{2}_{3i}$, \#148, $Z=2$ in rhombohedral and $Z=6$ in hexagonal cell).
The structure is formed by altering layers of corner sharing \ch{CoO6}\xspace and \ch{TiO6}\xspace octaherda, alternately stacked along the $c$ axis in the hexagonal settings as shown in Fig.~\ref{fig:CoTiO3_structure}~\cite{newnham1964crystal}. 
The lattice parameters at room temperature are $a=5.4846$\,\AA{}, $\theta=\ang{55.01}$ in rhombohedral setting and $a=b=5.056$, $c=13.91$\,\AA{} in hexagonal setting, as shown in Figs.~\ref{fig:CoTiO3_structure}a and b, respectively~\cite{newnham1964crystal,jacob2010role,harada2016magnetodielectric,balbashov2017electric}.
The hexagonal setting will be used throughout the paper below.
The unit cell contains 30 ions occupying the Wyckoff positions $6c$ (0, 0, 0.3551) for \ch{Co^{2+}}\xspace, $6c$ (0, 0, 0.1456) for \ch{Ti^{4+}}\xspace, and $18f$ (0.3162, 0.0209, 0.2459) for \ch{O^{2-}}\xspace as shown in Fig.~\ref{fig:CoTiO3_structure}b.
Below the N{\'e}el temperature $T_{N}=38$\,K, the $S=\frac{3}{2}$ spins of \ch{Co^{2+}}\xspace ($3d^7$) ions, form ferromagnetic honeycomb $ab$ planes which are antiferromagnetically coupled along the $c$ axis~\cite{newnham1964crystal,yuan2020dirac}, as can be seen in Fig.~\ref{fig:CoTiO3_structure}b.

The group-theoretical analysis for the ilmenite \CTO predicts 20 phonons~\cite{kroumova2003bilbao}
\begin{equation}
\label{eq:group_irrep_total}
\Gamma_{\textrm{total}} = 5 A_g \oplus 5 E_g \oplus 5 A_u \oplus 5 E_u,
\end{equation}
among which there are 2 acoustic $\Gamma_{\textrm{acoustic}}  =  A_u \oplus E_{u}$, 8 infrared active $\Gamma_{\textrm{IR}}  = 4 A_u \oplus 4 E_{u}$, and 10 Raman active $\Gamma_{\textrm{Raman}}  = 5 A_g \oplus 5 E_g$ phonons, where $A$ and $E$ are nondegenerate and doubly degenerate modes, respectively.

\subsection{Far-infrared reflectivity}

\begin{figure*}
\centering
\includegraphics[width=1\textwidth]{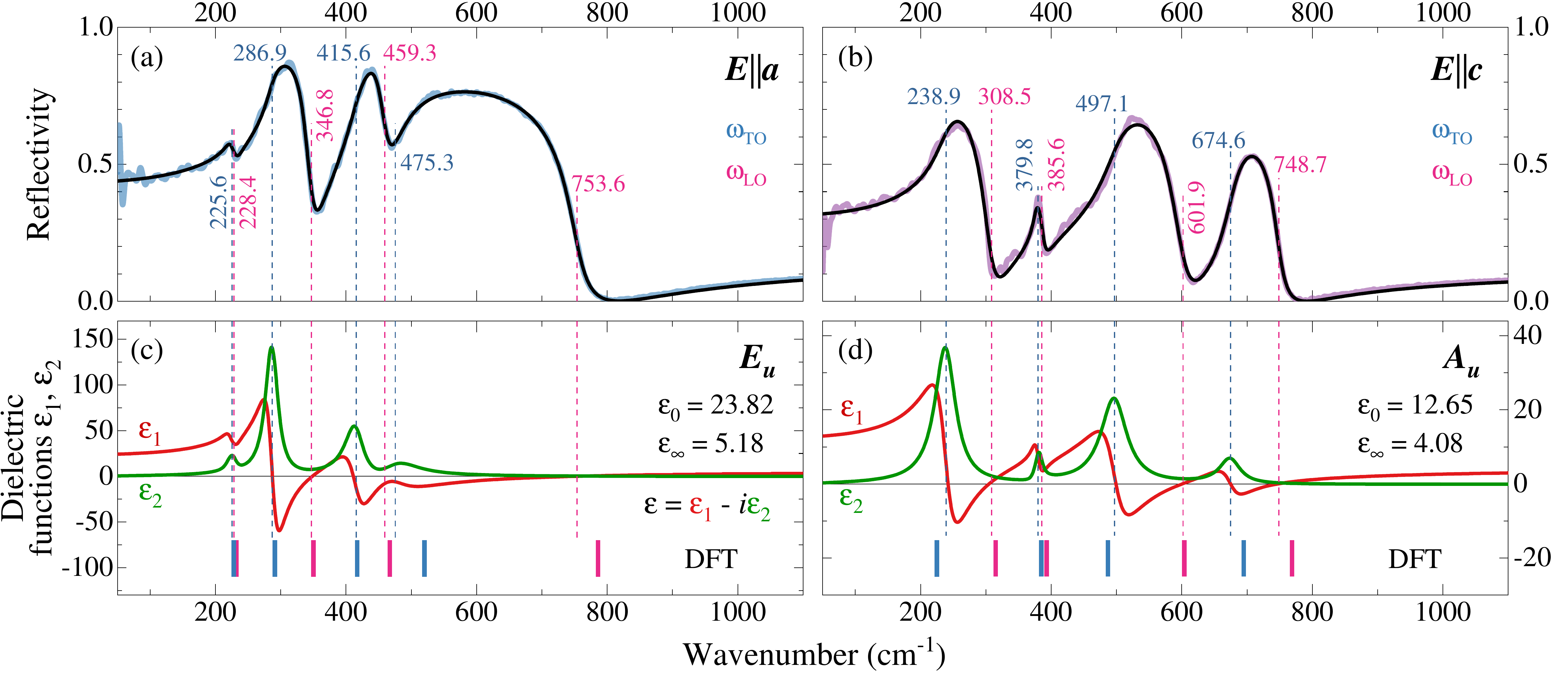}
\caption{\label{fig:reflectivity_spectra}
Far-infrared reflectivity spectra of \CTO at room temperature with the electric field of light $\bm{E}$ polarized along (a)~the $a$ axis and (b)~the $c$ axis.
The black lines are results of fits based on a generalized oscillator model according to Eq.~\eqref{eq:epsilon_TOLO}.
Spectra of real $\varepsilon_{1}$ and imaginary $\varepsilon_{2}$ parts of the complex dielectric function $\varepsilon$ for (c)~$A_{u}$ and (d)~$E_{u}$ phonons, corresponding to the fits.
Blue and magenta vertical dashed lines correspond to $\omega_{\textrm{TO}}$ and $\omega_{\textrm{LO}}$ phonon frequencies obtained from the fit, respectively.
The experimental values of the static $\varepsilon_{0}$ and high-frequency $\varepsilon_{\infty}$ dielectric permittivities are given.
Sticks in the lower parts of the (c)~and (d)~panels present the calculated phonon frequencies.
}
\end{figure*}

Figure~\ref{fig:reflectivity_spectra} shows the far-infrared reflectivity spectra of the ilmenite \CTO with polarization of light along the hexagonal axes at room temperature.
These reflectivity spectra are very similar to observed in the isostructural natural \ch{FeTiO3} crystal~\cite{hofmeister1993ir}. 
Four well-resolved bands in accordance with the symmetry prediction were detected for the both polarizations.
Reflectivity spectra were analyzed using the generalized oscillator model of complex dielectric function~\cite{gervais1974anharmonicity}
\begin{equation}
\label{eq:epsilon_TOLO}
\varepsilon(\omega) = \varepsilon_{1}(\omega) - i\varepsilon_{2}(\omega) = \varepsilon_{\infty}\prod\limits_{j}^{N}\frac{{\omega^{2}_{j\textrm{LO}}} - {\omega}^2 + i\gamma_{j\textrm{LO}}\omega}{{\omega^{2}_{j\textrm{TO}}} - {\omega}^2 + i\gamma_{j\textrm{TO}}\omega},
\end{equation}
where $\varepsilon_{\infty}$ is the high-frequency dielectric permittivity due to the electronic polarization contribution, $\omega_{j\textrm{LO}}$, $\omega_{j\textrm{TO}}$, $\gamma_{j\textrm{LO}}$ and $\gamma_{j\textrm{TO}}$ correspond to longitudinal $\textrm{LO}$ and transverse $\textrm{TO}$ frequencies and dampings of $j$th infrared active phonon, respectively.
$N$ is the number of polar phonons.
At near normal incidence, the reflectivity $R$ is related to the complex dielectric function by the Fresnel equation~\cite{born2013principles}
\begin{equation}
\label{eq:reflectivity}
R(\omega) = \Bigl|\frac{\sqrt{\varepsilon(\omega)} - 1}{\sqrt{\varepsilon(\omega)} + 1}\Bigr|^2.
\end{equation}
Results of fits of experimental data using Eqs.~\eqref{eq:epsilon_TOLO} and~\eqref{eq:reflectivity} are shown by black curves in Figs.~\ref{fig:reflectivity_spectra}a and~b.
There is a good agreement between experimental data and fits.
The spectra of real $\varepsilon_{1}$ and imaginary $\varepsilon_{2}$ parts of complex dielectric function $\varepsilon = \varepsilon_{1} - i\varepsilon_{2}$ calculated using obtained fit parameters are shown in Figs.~\ref{fig:reflectivity_spectra}c and~d.

\begin{table*}[width=1.2\linewidth,cols=7,pos=h]
    \caption{Parameters of the IR active phonons in \CTO at room temperature:
    frequencies $\omega_{j}$ (cm$^{-1}$), dampings $\gamma_{j}$ (cm$^{-1}$), and dielectric strengths $\Delta\varepsilon_{j}$.
    The results of DFT calculations are presented in parenthesis.} \label{tab:IR_phonon}
        \begin{tabular*}{\tblwidth}{@{} CCCCCCC@{} }
            \toprule
            \textbf{Sym.} & $\bm j$ & $\bm \omega_{j\textrm{TO}}$ & $\bm \gamma_{j\textrm{TO}}$ & $\bm \omega_{j\textrm{LO}}$ & $\bm \gamma_{j\textrm{LO}}$ & $\bm \Delta\varepsilon_{j}$\\
             \midrule
             \multirow{4}{*}{$E_{u}$}
            & 1 & 225.6 (228) & 12.3 & 228.4 (232) & 11.8 &  1.11 (1.39) \\
            & 2 & 286.9 (291) & 20.7 & 346.8 (350) & 23.4 & 11.03 (9.3) \\
            & 3 & 415.6 (417) & 33.4 & 459.3 (467) & 27.3 &  4.9  (3.41) \\
            & 4 & 475.3 (520) & 66.5 & 753.6 (786) & 34.3 &  1.61 (2.83) \\
            \cline{3-7}
            & & \multicolumn{5}{c}{$\varepsilon^{\textrm{opt}}_{0}=23.83$ (22.07) $ \qquad$ $\varepsilon_{\infty}=5.18$ (5.14)} \\
            \midrule
            \multirow{4}{*}{$A_{u}$}
            & 1 & 238.9 (225) & 36.9 & 308.5 (315) & 28.7 & 5.82 (9.16) \\
            & 2 & 379.8 (385) &  8.5 & 385.6 (393) & 13.4 & 0.2  (0.29) \\
            & 3 & 497.1 (487) & 46.6 & 601.9 (604) & 38.3 & 2.19 (2.27) \\
            & 4 & 674.6 (695) & 36.7 & 748.7 (769) & 20.2 & 0.38 (0.4) \\
            \cline{3-7}
            & &\multicolumn{5}{c}{$\varepsilon^{\textrm{opt}}_{0}=12.67$ (16.38) $ \qquad$ $\varepsilon_{\infty}=4.08$ (4.26)} \\
            \bottomrule
        \end{tabular*}
\end{table*}

Parameters of all predicted $4 A_{u}$ and $4 E_{u}$ polar phonons were reliably determined from the fits of experimantal data and are listed in Table~\ref{tab:IR_phonon}.
There is a qualitative agreement between experimental and literature phonon frequencies received from powder \CTO sample~\cite{baran1978ir}.
Obtained TO and LO phonon frequencies allow us to calculate the dielectric strength $\Delta\varepsilon_{j}$ of a $j$th phonon using an expression~\cite{gervais1983long}
\begin{equation}
\label{eq:oscillator_strength_TOLO}
\Delta\varepsilon_{j}  =  \frac{\varepsilon_{\infty}}{{\omega^{2}_{j\textrm{TO}}}}\frac{\prod\limits_{k}^{N}{\omega^{2}_{k\textrm{LO}}}-{\omega^{2}_{j\textrm{TO}}}}{\prod\limits_{k\neq{}j}^{N}{\omega^{2}_{k\textrm{TO}}}-{\omega^{2}_{j\textrm{TO}}}}.
\end{equation}
The optical static dielectric permittivity $\varepsilon^{\textrm{opt}}_{0}$ can be obtained by adding the dielectric strengths over all polar phonons according to expression
\begin{equation}
\label{eq:optic_epsilon}
\varepsilon^{\textrm{opt}}_{0} = \varepsilon_{\infty} + \sum_{j}^{N}\Delta\varepsilon_{j}.
\end{equation}
Values of the $\Delta\varepsilon_{j}$ for all polar phonons and $\varepsilon^{\textrm{opt}}_{0}$ are listed in Table~\ref{tab:IR_phonon}. 
It is worth noting that, in \CTO the values of $\varepsilon^{\textrm{opt}}_{0}$ along the $a$ and $c$ axes differ by about 2 times indicating strong anisotropy of the dielectric permittivity.
Moreover, the high-frequency dielectric permittivity $\varepsilon_{\infty}$ is also significantly different for the main crystallographic axes, as shown in Figs.~\ref{fig:reflectivity_spectra}a and~b.

We performed the DFT calculations of lattice dynamics at the Brillouin zone center of the ilmenite \CTO.
The obtained TO and LO frequencies of the polar phonons are given in parenthesis in Table~\ref{tab:IR_phonon} and are shown by sticks in Figs.~\ref{fig:reflectivity_spectra}c and~d.
The experimental phonon frequencies are slightly less than calculated due to the anharmonic effects appearing at room temperature~\cite{balkanski1983anharmonic}.
Nevertheless, there is a good agreement between experimental and calculated results.
Furthermore, the strong anisotropy of $\varepsilon^{\textrm{opt}}_{0}$ and $\varepsilon_{\infty}$ dielectric permittivity were also well reproduced in our computations.
We add that a strong dielectric anisotropy has been theoretically predicted for isostructural \ch{FeTiO3}\xspace~\cite{FeTiO3_database} and \ch{MgTiO3}\xspace~\cite{MgTiO3_database} and experimentally observed in \ch{CdTiO3}\xspace~\cite{rodrigues2020unveiling} ilmenite crystals. 

\begin{figure*}
\centering
\includegraphics[width=1\textwidth]{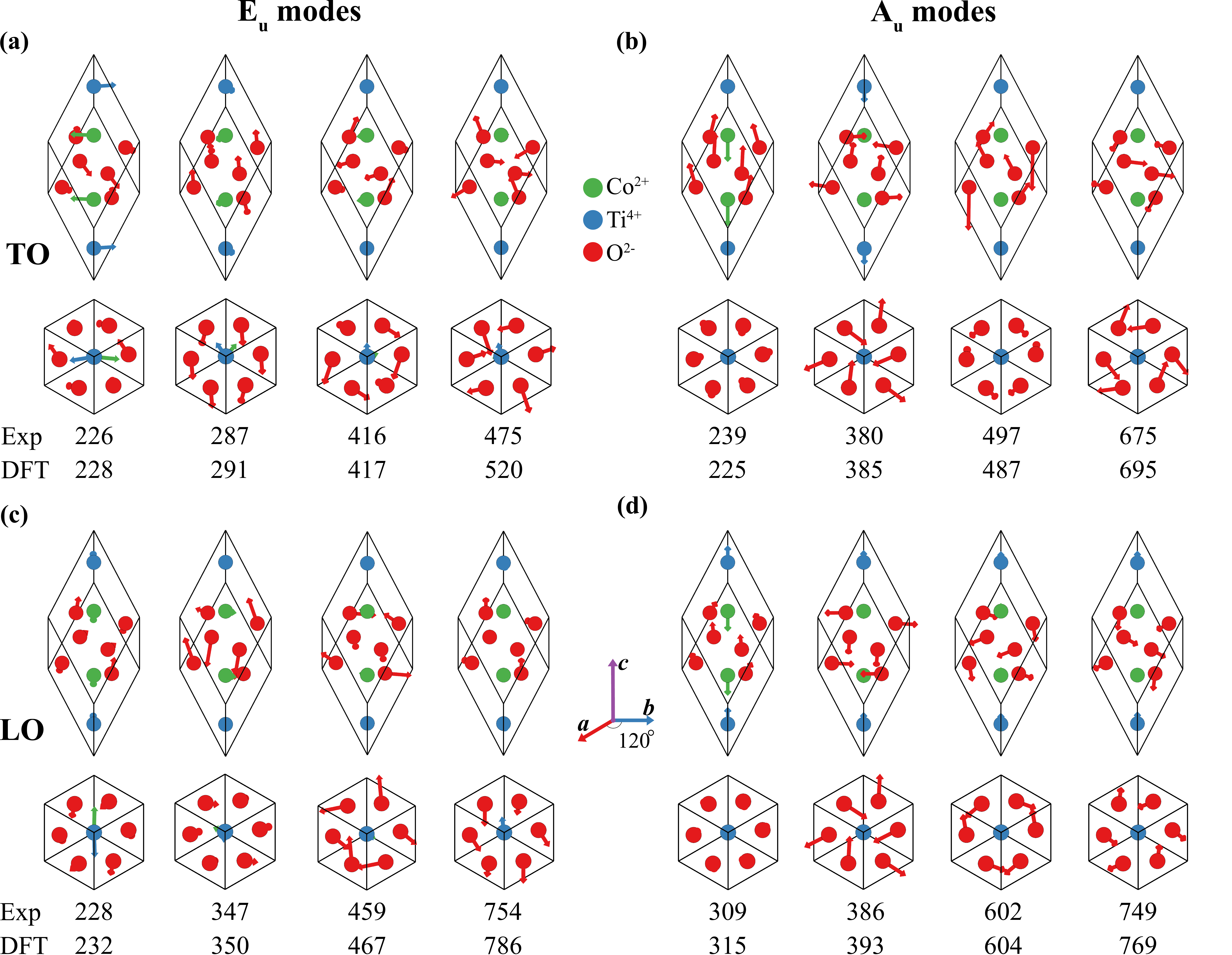}
\caption{\label{fig:IR_phonons}
Schematic representation of ionic displacements in the Brillouin zone center in \CTO for polar (a,c)~$4E_{u}$ and (b,d)~$4A_{u}$ TO and LO phonons the in two projections according to the DFT calculations.
For clarity, the crystallographic axes are shown for the hexagonal setting.
The relative amplitudes of vibrations are illustrated by arrows.
The experimental and calculated values of phonon frequencies (cm$^{-1}$) are given.
}
\end{figure*}

The DFT calculations also provide the normal mode eigenvectors for ions.
Figure~\ref{fig:IR_phonons} shows obtained ionic displacement patterns for all infrared active $4E_{u}$ and $4A_{u}$ phonons in ilmenite \CTO.
It was found that the TO and LO modes of the $E_{u}$ symmetry are related to the $ab$ plane vibrations of \ch{Co} and \ch{Ti} ions, as can be seen in Figs.~\ref{fig:IR_phonons}a and c, respectively.
Whereas the $A_{u}$ phonons mainly represent the vibrations of \ch{Co} and \ch{Ti} ions along the $c$ axis for both TO and LO modes (see Figs.~\ref{fig:IR_phonons}b and d).
For all infrared active phonons the movements of oxygen anions correspond to the asymmetric breathing.
It is worth noting that the calculated pattern of vibrational eigenvectors in ilmenite \CTO are close to the data for isostructural \ch{CdTiO3}~\cite{rodrigues2020unveiling} and \ch{MgTiO3}~\cite{wang2008far}.

\begin{figure}
\centering
\includegraphics[width=\columnwidth]{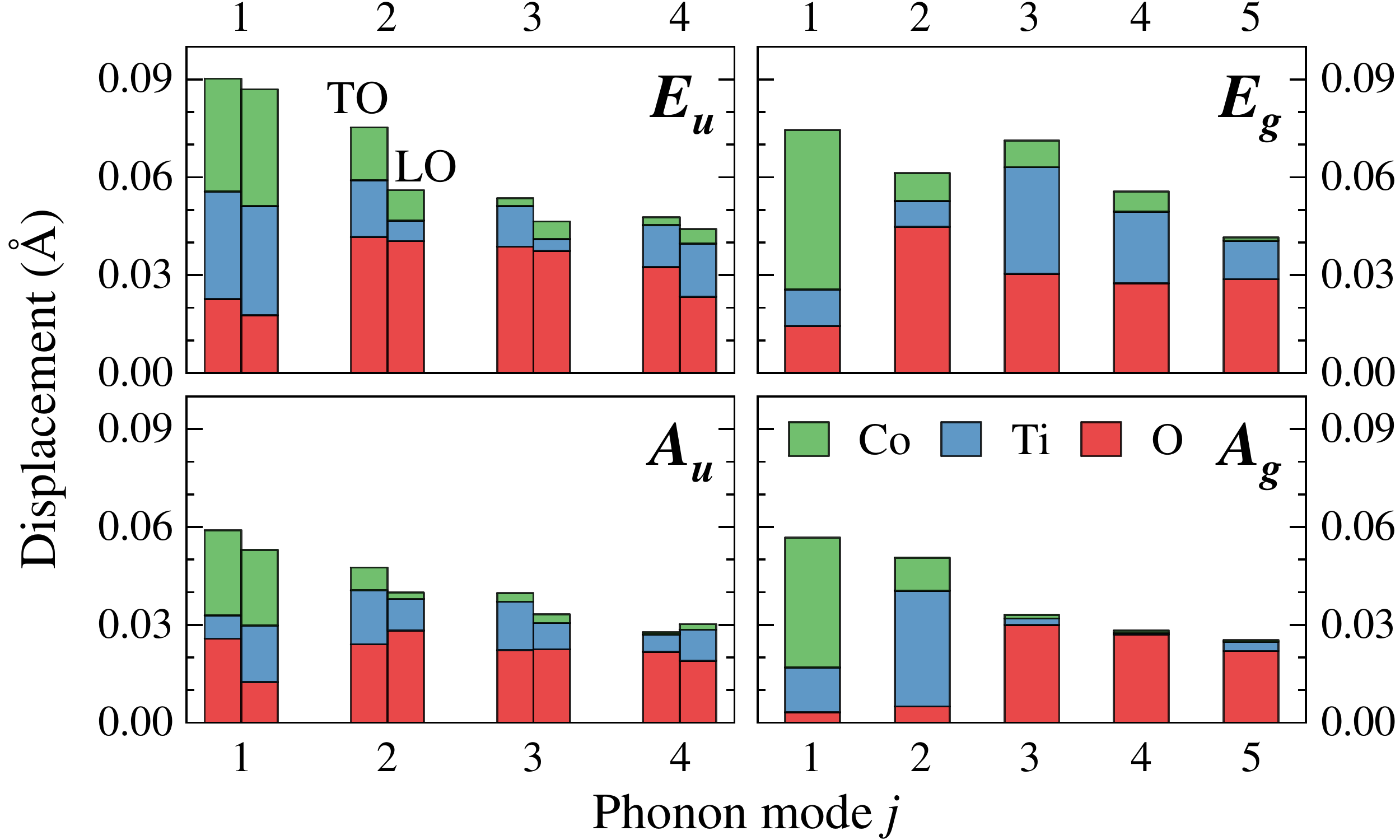}
\caption{\label{fig:displacement}
Average displacements of ions for infrared ($4E_{u} \oplus 4A_{u}$) and Raman ($5E_{g} \oplus 5A_{g}$) active phonon modes in ilmenite \CTO according to the DFT calculations.
}
\end{figure}

Figure~\ref{fig:displacement} shows as stacked bar graphs the calculated average displacement of ions for $4E_{u} \oplus 4A_{u}$ infrared (left panels) and  $5E_{g} \oplus 5A_{g}$ Raman (right panels) active phonon modes.
For $E_{u}$ and $A_{u}$ modes, the displacements of \ch{Co} ions are significant only for low-frequency phonons as can be seen on the left panel in Fig.~\ref{fig:displacement}.
Apparently, this is due to the fact that \ch{Co} is the heaviest ion in this crystal.
The average amplitudes of \ch{Ti} and \ch{O} ion vibrations are depend on phonon modes more weakly.
For some phonons, there is a significant difference in average displacements between TO and LO modes.
Thus, for $E_{u}$ phonons with $j=2$ and $3$ the amplitudes of \ch{Ti} displacements for the TO modes are much larger then for the LO modes.

\subsection{Raman scattering}

\begin{figure*}
    \centering
    \includegraphics[width=1\textwidth]{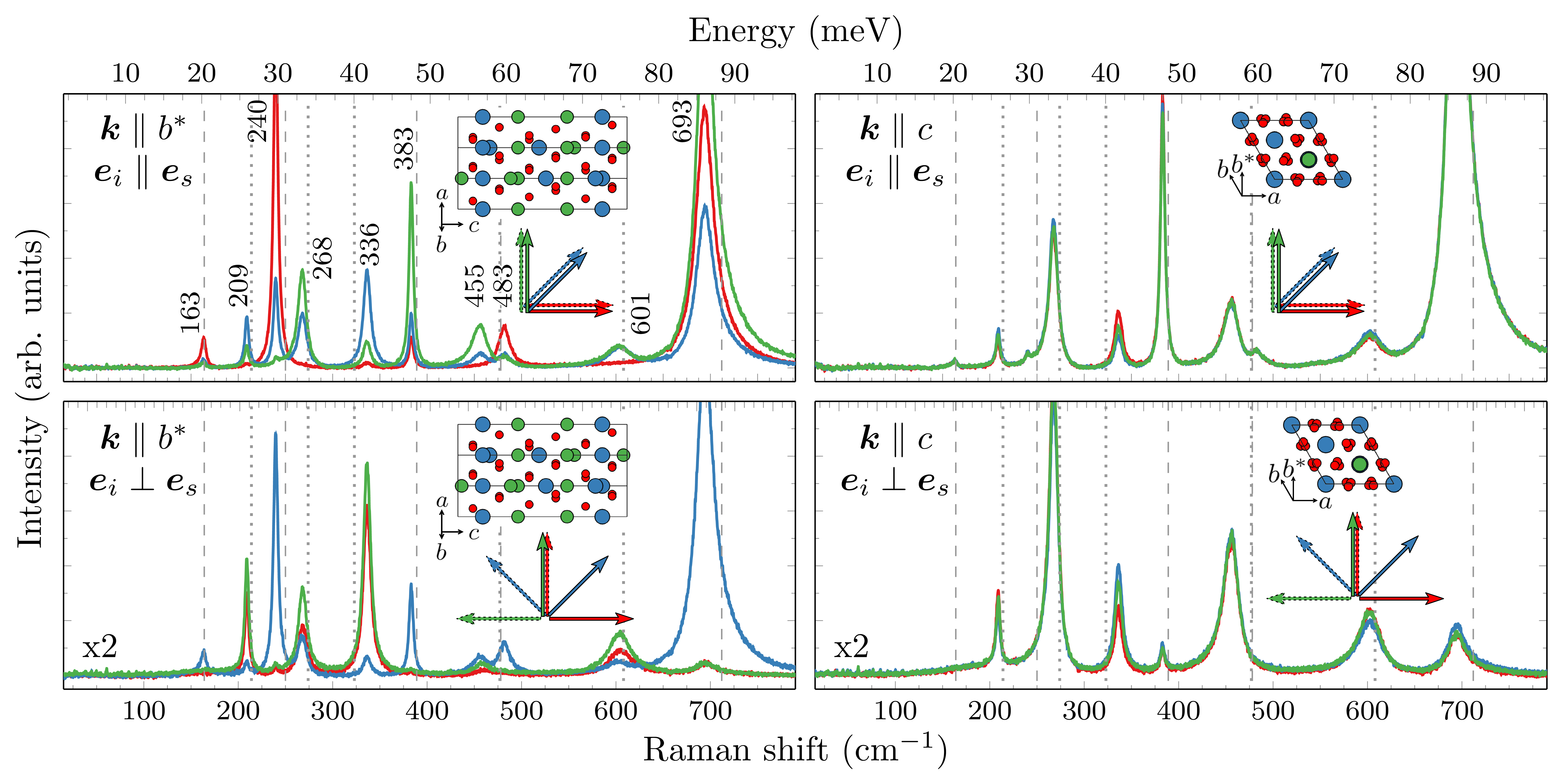}
    \caption{
        \label{fig:raman_spectra}
        Raman scattering spectra of \CTO at room temperature.
        Linear background and weak quasielastic tail are subtracted.
        Insets shows crystal structure projections in the $ac$- and $ab$-planes in left and rights plots, respectively.
        Arrows represent the polarization of incident ($\bm{e}_i$, solid outline) and scattered ($\bm{e}_s$, dashed outline) light with respect to the crystallographic axes.
        The intensity of bottom plots is scaled by $\times2$ to compensate weaker scattering in crossed polarizations.
        Vertical dashed and dotted lines correspond to calculated frequencies for $A_g$ and $E_g$ phonons, respectively.
        Numbers represent frequencies of the corresponding phonon modes.
    }
\end{figure*}

Raman scattering is a powerful technique allowing observation of different types of even excitations.
As suggested in Ref.~\cite{yuan2020dirac} Raman technique could be applied for observation of bulk and surface magnons deep in the antiferromagnetic phase of \CTO, however the reliable assignment of the bulk lattice modes is required beforehand.
There are a few reports on Raman scattering studies of \CTO, however they all were done on powder~\cite{baraton1994vibrational} and polycrystalline samples~\cite{fujioka2016raman}, which do not allow definitive determination of the phonon symmetries.

\CTO is a challenging crystal for Raman measurements due to its high optical density leading to poor scattering efficiency and unavoidable overheating.
These restrictions can be avoided by choosing appropriate excitation line corresponding to one of the transparency windows.
Two sets of spectra were recorded in the $ac$ and $ab$ planes, respectively, with the light linearly polarized along principal crystal axes, as well as for some intermediate directions, as shown in insets in Fig.~\ref{fig:raman_spectra}. 
As expected, the $A_g$ modes are isotropic in the $ab$ plane, while showing considerable intensity difference in $ac$ plane, according to scattering tensors~\cref{eq:raman_tensors} suggesting considerable difference in the $a$ and $b$ tensor elements for all modes.

\begin{equation}
    \label{eq:raman_tensors}
    \begin{gathered}
        A_g =
        \begin{pmatrix}
             a& 0 & 0 \\ 
             0& a & 0 \\ 
             0& 0 & b
        \end{pmatrix}
        ,\quad
        \textrm{E}_{g} = 
        \begin{pmatrix}
             c&  d & e \\ 
             d& -c & f \\ 
             e&  f & 0
        \end{pmatrix}
    \end{gathered}
\end{equation}

\begin{table}[width=0.9\linewidth,cols=4,pos=h]
    \caption{Phonon frequencies (\cm) in comparison with results of DFT calculations, shown in parentheses.
    Full width at half maxima (FWHM, \cm) measured for polarization where the corresponding mode is strongest.}\label{tab:phonon_raman}
    \begin{tabular*}{\tblwidth}{@{} CCCC@{} }
        \toprule
        \textbf{Symmetry} & \textbf{Mode} & \textbf{Frequency}  & \textbf{FWHM}\\
        \midrule
        \multirow{5}{*}{$E_{g}$}
            &   1   &   209 (214) &   5.7     \\
            &   2   &   268 (273) &   11.2    \\
            &   3   &   336 (323) &   10.3    \\
            &   4   &   455 (474) &   20.9    \\
            &   5   &   601 (608) &   32.0    \\
            \midrule
        \multirow{5}{*}{$A_{g}$}
            &   1   &   163 (164) &   6.8     \\
            &   2   &   240 (250) &   6.7     \\
            &   3   &   383 (389) &   6.4     \\
            &   4   &   483 (477) &   15.8    \\
            &   5   &   693 (712) &   25.0    \\
        \bottomrule
    \end{tabular*}
\end{table}

Small leakage of phonon modes is expected due to the light depolarization on optical elements and minor misalignment of the sample axes with respect to the light polarization.
All even Raman-active modes predicted by symmetry analysis (see~\cref{eq:group_irrep_total}) were observed in the range 150--800\,\cm with their individual frequencies summarized in~\cref{tab:phonon_raman}.
Generally, the frequencies of these modes are in a good accordance with the results of DFT calculations, however they are slightly softer due to anharmonic effects at finite, room in our experiments, temperature, similar to the case of the polar phonons.

\begin{figure*}
\centering
\includegraphics[width=1\textwidth]{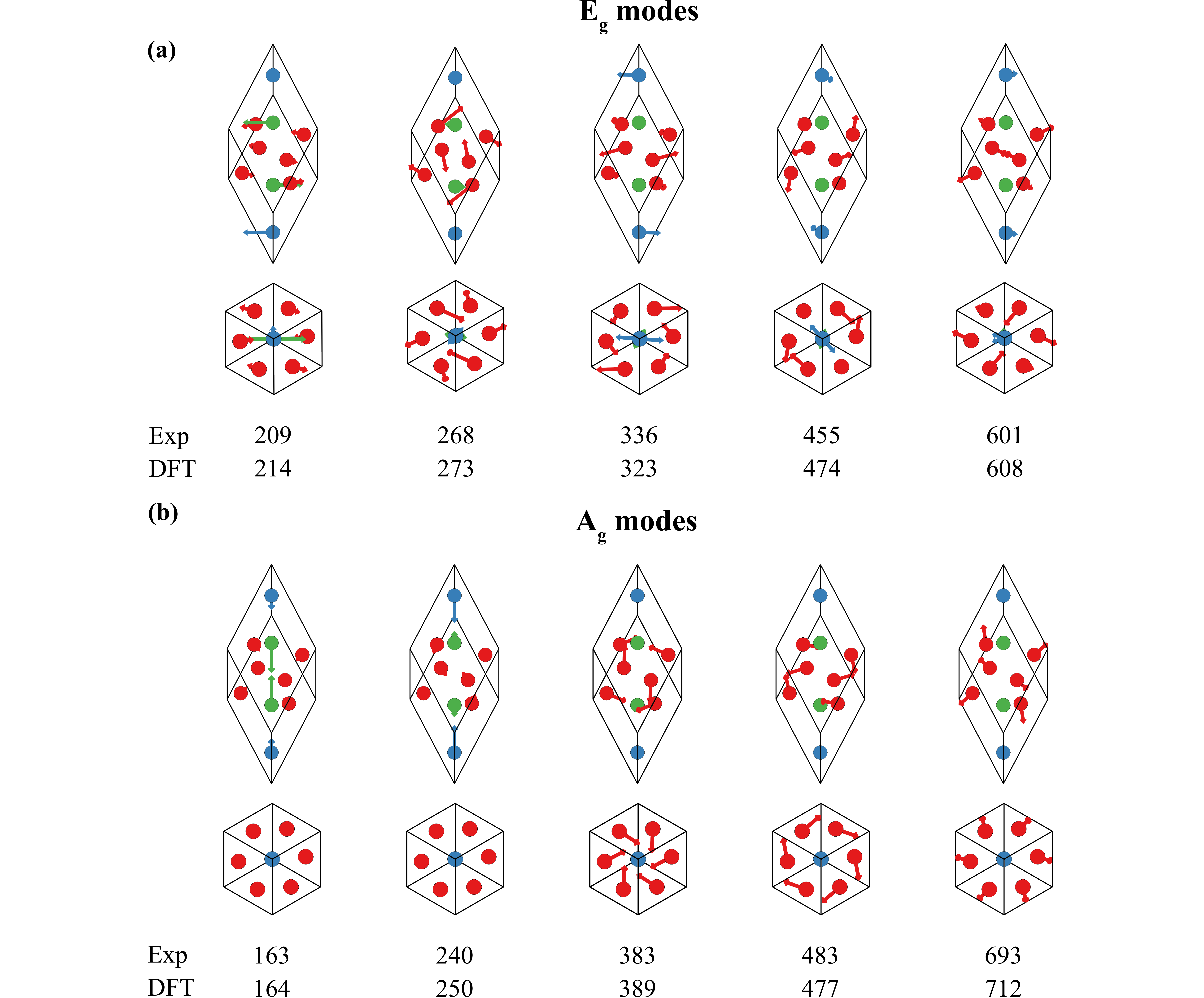}
\caption{\label{fig:Raman_phonons}
Schematic representation of ionic displacements in the two projections in \CTO for Raman active (a)~$5E_{g}$ and (b)~$5A_{g}$ phonons in the Brillouin zone center according to the DFT calculations.
The relative amplitudes of vibrations are shown by arrows.
The experimental and calculated values of phonon frequencies (cm$^{-1}$) are given.
}
\end{figure*}

Notably, the phonon of the highest frequency 693\,\cm has strongly asymmetric shape, and can by more or less equally good described by two symmetric or one asymmetric Voigt profiles.
In Ref.~\cite{fujioka2016raman} this asummetric mode was interpreted as the two separate ones.
The shape of this mode is similar in all polarizations, and there is no any additional modes at this energy according to the DFT calculations.
It seems unlikely that there is an additional mode with frequency $\approx 711$~\cm, e.g., created by two-phonon scattering processes having nearly the same Raman tensor elements responsible for the  693\,\cm excitation, and therefore we may conclude that its asymmetry is intrinsic.

The $b^*(c'c')\overline{b^*}$ geometry allows the simultaneous observation of all the even phonon modes with considerable intensity.
Having complete information of relative intensity of the modes, the polarized Raman scattering can be regarded as a supplementary tool for the easy and fast orientation of \CTO.

Figure~\ref{fig:Raman_phonons} shows the ionic displacement patterns for $5E_{g}$ and $5A_{g}$ Raman active modes in ilmenite \CTO which are similar to those for isostructural \ch{CdTiO3}~\cite{rodrigues2018first} and \ch{MgTiO3}~\cite{wang2008assignment}.
According to symmetry analysis, displacements of \ch{Co} and \ch{Ti} ions occur in the $ab$ plane for all $E_{g}$ modes, while for $A_{g}$ modes these ions move along the $c$ axis. 
For the lowest-frequency $E_{g}$ and $A_{g}$ phonons the opposite directional displacements of \ch{Co} ions are significant, as can be seen on the right panels in Fig.~\ref{fig:displacement}.
The average displacements of \ch{Ti} ions are predominant for the $A_{g}$ mode with $j=2$.
It should be noted that the $A_{g}$ modes with $j=1,2$ are characterized by relatively small average displacements of \ch{O} ions as shown in Figs.~\ref{fig:displacement} and~\ref{fig:Raman_phonons}b.
For other $A_{g}$ modes, the average displacements of \ch{Co} and \ch{Ti} ions are relatively small.
The displacements of \ch{Ti} and \ch{O} ions for $E_{g}$ phonons are characterized by a non-monotonic dependence on vibrational modes as shown in Fig.~\ref{fig:displacement}.

\cite{rodrigues2018first}

\subsection{Dynamical modulation of the superexchange interactions by phonons}

\begin{figure}
\centering
\includegraphics[width=\columnwidth]{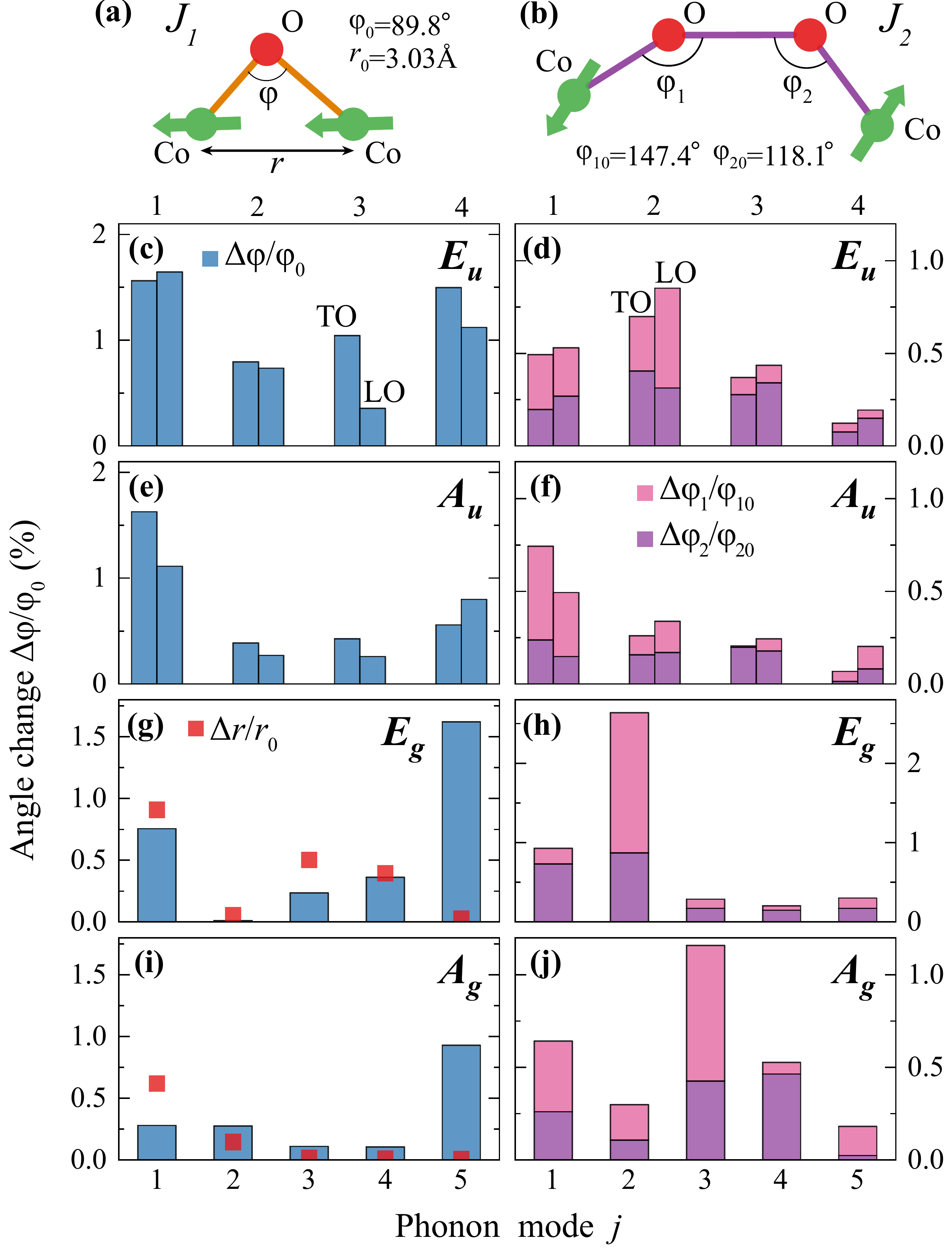}
\caption{\label{fig:CoTiO3_J_modulation}
The calculated bond angle changes $\Delta\phi/\phi_{0}$ of the superexchange (a)~$J_{1}$ and (b)~$J_{2}$ interactions due to the ionic displacements for (c,d)~$E_{u}$, (e,f)~$A_{u}$, (g,h) $E_{g}$ and (i,j) $A_{g}$ optical phonons, respectively.
Additionally, Raman active $E_{g}$ and $A_{g}$ phonons dynamically change distances  $\Delta{r}/{r_{0}}$ of the direct exchange between nearest neighbor \ch{Co} ions as shown by the red squares.
Arrows represent the spin directions.
The average values of angles and distance of superexchange paths for both $J_{1}$ and $J_{2}$ exchange interactions are given.
}
\end{figure}

The spin-phonon coupling give rise a numerous intriguing physical phenomena in magnetic materials~\cite{son2019unconventional}.
This coupling manifest itself in shifts of the phonon frequencies due to magnetic ordering and originates from the modulation of the superexchange integrals by lattice vibrations.
It is known that the superexchange interaction strongly depends on the wave function overlap of the electron clouds and, therefore, is extremely sensitive to the relative distances and bond angles between magnetic ions~\cite{granado1999magnetic,streltsov2017orbital}.
The DFT calculations allowed us to estimate the relative magnitude of the dynamical changes of the geometric superexchange paths caused by ionic displacements shown in Figs.~\ref{fig:IR_phonons} and~\ref{fig:Raman_phonons}. 
The magnetic structure of ilmenite \CTO can be adequately described by the two exchange integrals with the values $J_{1}=-4.5$\,meV and $J_{2}=0.6$\,meV~\cite{yuan2020dirac}.

The anisotropic ferromagnetic $J_{1}$ interaction within the $ab$ planes is formed by the direct cation-cation exchange coupling between the nearest-neighbor cations and the 90$^{\circ}$ cation-anion-cation superexchange coupling~\cite{goodenough1967theory} as shown in Figs.~\ref{fig:CoTiO3_structure}b and~\ref{fig:CoTiO3_J_modulation}a.
The displacements of ions for infrared active $E_{u}$ and $A_{u}$ phonons shown in Fig.~\ref{fig:IR_phonons} lead to a dynamic change of the superexchange \ch{Co}--\ch{O}--\ch{Co} bong angles $\Delta\phi/\phi_{0}$.
For Raman active $E_{g}$ and $A_{g}$ phonons the vibrations of ions (see Fig.~\ref{fig:Raman_phonons}), besides the bond angles, also lead to dynamical changes of the relative distance $\Delta{r}/r_{0}$ of the direct \ch{Co}--\ch{Co} exchange interaction.
The isotropic antiferromagnetic superexchange $J_{2}$ interaction occurs via next-nearest-neighbor coupling along the cation-anion-anion-cation path in the direction of the $c$ axis~\cite{goodenough1967theory} as shown in Figs.~\ref{fig:CoTiO3_structure}b and~\ref{fig:CoTiO3_J_modulation}b.
For the $J_{2}$ interaction all lattice vibrations dynamically change bond angles $\phi_{1}$ and $\phi_{2}$ of the \ch{Co}--\ch{O}--\ch{O}--\ch{Co} path only.

Figures~\ref{fig:CoTiO3_J_modulation}c--j present the calculated relative changes of angles (bars) and distance (squares) for superexchange $J_{1}$ and $J_{2}$ interactions for studied phonons.
Analysis of the obtained results allows us to predict the relative values of the phonon frequency shifts at antiferromagnetic ordering caused by the spin-phonon coupling.
Since the absolute value of $J_{1}$ is almost an order of magnitude larger than $J_{2}$, we took into account only the effects of the strongest exchange interaction on the phonon frequencies.
We predict that the spin-phonon coupling should manifest itself for polar $E_{u}$ and $A_{u}$ phonons proportional to the dynamical modulation of the $\phi$ angle (see Figs.~\ref{fig:CoTiO3_J_modulation}c and d).
Furthermore, for the $E_{u}$ with $j=3$ phonon the significant spin-phonon coupling is expected for $\omega_{\textrm{TO}}$, while for $\omega_{\textrm{LO}}$ this effect is expected to be much weaker.
One may expect that a change in the relative distance $r$ between \ch{Co}--\ch{Co} ions has a stronger effect on the $J_{1}$ interaction than the bond $\phi$ angle.
Thus, the largest frequency shifts are expected for the Raman active $E_{g}$ and $A_{g}$  phonons with frequencies 163, 209, 336 and 455\,cm$^{-1}$ due to dynamic modulation of the direct distance $\Delta{r}/r_{0}$ as shown in Figs.~\ref{fig:CoTiO3_J_modulation}g and i.
It should be noted that possible deviations between the predicted and experimentally observed phonon frequency shifts may be related to magnetoelastic coupling~\cite{dey2020magnetic}, strong anisotropy of $J_{1}$~\cite{yuan2020dirac} and effect of $J_{2}$ interactions.
Nevertheless, all these computational predictions require experimental confirmation.

\subsection{Dielectric spectroscopy}

\begin{figure*}
\centering
\includegraphics[width=1\textwidth]{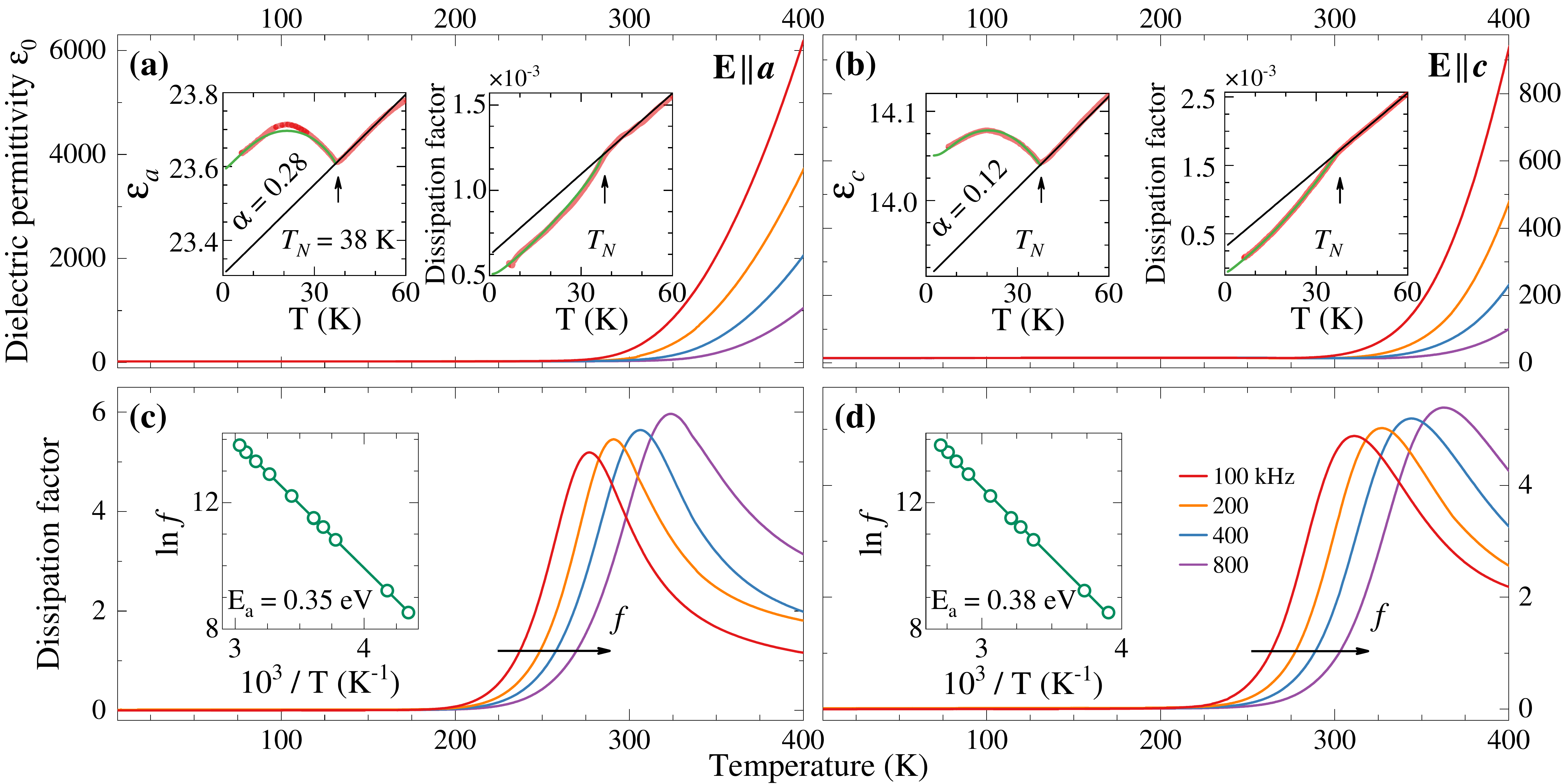}
\caption{\label{fig:dielectric}
Temperature dependencies of the (a), (b) dielectric permittivity and (c), (d) dissipation factor at the indicated frequencies in \CTO along the $a$ and $c$ axes, respectively.
Insets in (a) and (b) show changes of $\varepsilon_{0}(T)$ at 100\,kHz below $T_{N}$ due to antiferromagnetic ordering for relevant axes.
The black lines are the expected behavior of $\varepsilon_{0}^{\textrm{NM}}$ assuming absence of antiferromagnetic ordering.
The green lines are fits of the shifts due to spontaneous magnetodielectric effect according to Eq.~\eqref{eq:epsilon_MD}.
Arrhenius plots obtained from dissipation factor data are shown on insets in (c) and (d) for $a$ and $c$ axes, respectively.}
\end{figure*}

Temperature dependencies of the dielectric permittivity and dissipation factor in \CTO in the temperature range 5--400\,K at frequencies from 5\,kHz to 1\,MHz are shown in Fig.~\ref{fig:dielectric}.
The absolute values of the $\varepsilon_{0}$ were normalized to the corresponding quantities from far-infrared reflectivity. 
Significant increase of the dielectric permittivity with essential frequency dispersion is observed at high temperatures for both main axes, as can be seen in Figs.~\ref{fig:dielectric}a and~b.
Moreover, the dissipation factor demonstrate peaks in the same temperature range which are also frequency dependent and obey the Arrhenius law $f = f_{0} \exp{\cfrac{-E_{a}}{k_{B}T}}$ as shown on insets in Figs.~\ref{fig:dielectric}c and~d for $a$ and $d$ axes, respectively.
The obtained values of activation energy $E_{a}=0.35$ and $0.38$\,eV are close to the those for ceramic sample of \CTO from Ref.~\cite{acharya2016structural}. 
This relaxation process is apparently due to hopping conductivity~\cite{elliott1987ac} which was previously observed in the related inverse spinel \ch{Co2TiO4}~\cite{prosnikov2016lattice} and some other oxide crystals~\cite{kant2008optical,savinov2008dielectric,bamzai2011dielectric}.

The distinct anomaly in the temperature dependence of the dielectric permittivity is observed for both $a$ and $c$ axes at $T_{N}=38$\,K due to to the antiferromagnetic ordering, as shown on insets in Figs.~\ref{fig:dielectric}a and~b, respectively.
It is interesting to note that in \ch{MnTiO3}\xspace which are isostructural but with different magnetic structure the anomalies of the dielectric permittivity below $T_{N}$ were observed only in an external magnetic field~\cite{mufti2011magnetoelectric}.
The expected behavior of the dielectric permittivity $\varepsilon_{0}^{\textrm{NM}}$ assuming absence of antiferromagnetic ordering at low temperatures is shown by black lines on insets in Figs.~\ref{fig:dielectric}a and~b.
The deviation of $\varepsilon_{0}(T)$ from the expected behavior of $\varepsilon_{0}^{\textrm{NM}}(T)$ below $T_{N}$ due the spontaneous magnetodielectric effect was fitted by expression~\cite{katsufuji2001coupling}
\begin{equation}
\label{eq:epsilon_MD}
\Delta\varepsilon^{\textrm{MD}}(T) = \alpha \langle S_{i} \cdot{} S_{j} \rangle, 
\end{equation}
where $\alpha$ is a coefficient and $\langle S_{i} \cdot{} S_{j} \rangle$ is the spin-pair correlation function between nearest-neighbor spins, which was represented as the squared Brillouin function for simplicity~\cite{darby1967tables}.
There is a good agreement between experimental (red) and fit (green) lines on insets in Figs.~\ref{fig:dielectric}a and~b.  

Obtained values of spontaneous magnetodielectric effect accounts for about $\alpha=0.28$ and $0.12$ for the $a$ and $c$ axes, respectively.
It is worth noting that these values $\alpha$ are comparable in magnitude as those observed in polycrystalline ilmenites \CTO and \ch{NiTiO3}\xspace~\cite{harada2016magnetodielectric}.
However, in our experiments the spontaneous magnetodielectric effect is positive in \CTO, i.e., it leads to increase of the dielectric permittivity below $T_{N}$ for the both main crystallographic axes whereas in polycrystalline sample this effect was negative~\cite{harada2016magnetodielectric}.
Moreover, the temperature behavior of $\Delta\varepsilon^{\textrm{MD}}$ in single crystal of \CTO is similar to that observed in \ch{NiTiO3}\xspace with the same crystal and magnetic structures~\cite{harada2016magnetodielectric}.

The spontaneous magnetodielectric effect can be caused by discussed above the dynamical modulation of the exchange interaction by infrared active phonons~\cite{dubrovin2019lattice} and by direct crystal volume changes at magnetic ordering due to magnetoelastic coupling~\cite{xie2016magnetodielectric}.
It was suggested that the magnetodielectric effect in ilmenite \ch{NiTiO3}\xspace single crystal is mainly related to the magnetoelastic coupling~\cite{dey2020magnetic}.
Also, the spontaneous magnetostriction was experimentally observed in isostructural \ch{FeTiO3}\xspace \cite{charilaou2012large}.
It is interesting to note that in the related inverse spinel \ch{Co2TiO4} no anomaly in the temperature dependence of the dielectric permittivity at the ferrimagnetic phase transition was revealed~\cite{prosnikov2016lattice}.
In our opinion, the observed spontaneous magnetodielectric effect in ilmenite \CTO has a primarily magnetoelastic origin whereas the effect of spin-phonon coupling is less pronounced.
This suggestion is confirmed by the fact that not only the dielectric permittivity, but also the dissipation factor shows changes due to the antiferromagnetic ordering as can be seen on insets in Figs.~\ref{fig:dielectric}c and~d.
Further experimental studies of the lattice dynamics at low temperatures of ilmenite \CTO could reveal more deep insight into the origin of the magnetodielectric coupling.

\section{Conclusions}

In summary, we report the results of detailed lattice dynamics studies of cobalt titanite \CTO single crystal with ilmenite structure using complementary dielectric, far-infrared, and Raman polarized spectroscopic techniques.
Obtained experimental results, supplemented by the DFT calculations, allowed us to reliably identify the frequencies and symmetries of all predicted infrared and Raman active phonons.
It was shown that the antiferromagnetic ordering is accompanied by changes the low-frequency dielectric permittivity in zero external magnetic field due to the spontaneous magnetodielectric effect.
We believe that our results will stimulate further research in deep of \CTO and other ilmenite-type crystals.

\section*{Declaration of competing interest}

The authors declare that they have no known competing financial interests or personal relationships that could have appeared to influence the work reported in this paper.

\section*{Acknowledgements}

We thank D.\,A.\,Andronikova and M.\,P.\,Scheglov for the help with the X-ray orientation of the single crystal and K.\,N.\,Boldyrev for scientific discussion.
M.\,A.\,P. acknowledges fruitful discussions with Beatrice\,T.\,Crow.
N.\,V.\,S., R.\,M.\,D. and R.\,V.\,P. are grateful to Russian Science Foundation [project number 16-12-10456] for financial support.
Besides, the authors would like to thank the HFML-RU/FOM, a member of the European Magnetic Field Laboratory (EMFL) for support of the Raman spectroscopy study.

\printcredits

\bibliographystyle{elsarticle-num}

\bibliography{biblio.bib}

\end{document}